\documentclass[%
 reprint,
 superscriptaddress,
 showkeys,
nofootinbib,
 amsmath,amssymb,
 aps,
prc,
floatfix
]{revtex4-2}
\usepackage{longtable}
\usepackage{pgfplotstable}
\usepackage{graphicx}
\usepackage{dcolumn}
\usepackage{bm}
\usepackage{float}
\usepackage{xcolor}
\usepackage{hyperref}
\usepackage{tabularx}
\usepackage{booktabs}
\usepackage{bigdelim}
\usepackage{afterpage}

\DeclareUnicodeCharacter{2212}{-}

\definecolor{lightgray}{rgb}{0.77, 0.76, 0.82}

\newcolumntype{H}{>{\setbox0=\hbox\bgroup}c<{\egroup}@{}}
\newcolumntype{L}[1]{>{\raggedright\let\newline\\\arraybackslash\hspace{0pt}}m{#1}}
\newcolumntype{C}[1]{>{\centering\let\newline\\\arraybackslash\hspace{0pt}}m{#1}}
\newcolumntype{R}[1]{>{\raggedleft\let\newline\\\arraybackslash\hspace{0pt}}m{#1}}

\begin{document}


\title{Electric dipole strength in $sd$-shell nuclei from small-angle proton scattering}

\author{R.~W.~Fearick}
\affiliation{Department of Physics, University of Cape Town, Rondebosch 7700, South Africa}

\author{O.~Le~Noan}
\affiliation{Universit\'e de Strasbourg, IPHC, 23 rue du Loess, 67037 Strasbourg, France}
\affiliation{CNRS, UMR7178, 67037 Strasbourg, France}

\author{H.~Matsubara}
\affiliation{Research Center for Nuclear Physics, Osaka University, Ibaraki, Osaka 567-0047, Japan}
\affiliation{School of Medical Sciences, Fujita Health University, Aichi 470-1192, Japan}

\author{P.~von Neumann-Cosel}\email{email:vnc@ikp.tu-darmstadt.de}
\affiliation{Institut f\"{u}r Kernphysik, Technische Universit{\"a}t Darmstadt, D-64289 Darmstadt, Germany}

\author{K.~Sieja}
\affiliation{Universit\'e de Strasbourg, IPHC, 23 rue du Loess, 67037 Strasbourg, France}
\affiliation{CNRS, UMR7178, 67037 Strasbourg, France}

\author{A.~Tamii}
\affiliation{Research Center for Nuclear Physics, Osaka University, Ibaraki, Osaka 567-0047, Japan}

\date{\today}

\begin{abstract}
\begin{description}

\item[Background]
Total photoabsorption cross sections are important for an understanding of nuclear structure and for many applications, in particular in astrophysics.
The majority of data stems from measurements of neutron emission after photoexcitation, which in heavy nuclei represent a good approximation to the total cross sections.
However, this does not hold for light nuclei, where charged-particle emission thresholds are typically lower in energy. Accordingly, data in light nuclei are scarce.

\item[Purpose]
The present work reports new total photoabsorption cross sections for the $N=Z$ nuclei in the $sd$-shell $^{20}$Ne, $^{24}$Mg, $^{28}$Si, $^{32}$S, $^{36}$Ar, and for $^{26}$Mg. 
The results are compared to predictions of a data-driven artificial neural network application and to configuration-interaction shell-model calculations covering the excitation energy region of the isovector giant dipole resonance.

\item[Methods]

Double-differential cross sections of the  $(p,p^\prime)$ reaction at an incident proton energy of 295 MeV have been measured between $0^\circ$ and $14^\circ$.
The angular distributions of the $E1$ parts due to Coulomb excitation have been extracted with a multipole decomposition analysis for excitation energies 12 to 24 MeV and converted to equivalent photoabsorption cross sections with the virtual photon method. 

\item[Results]
Reasonable agreement of the photoabsorption cross sections with previous experiments is found for $^{24}$Mg and $^{28}$Si, while the present results diverge for $^{32}$S.
For the first time, data are presented for $^{20}$Ne, $^{26}$Mg and $^{36}$Ar.
Configuration-interaction shell-model calculations provide an overall satisfactory description of the fragmented $E1$ strength distributions. 
The same holds for absolute cross sections except for $^{26}$Mg and $^{36}$Ar, where the experimental results significantly exceed the expected exhaustion of the Thomas-Reiche-Kuhn energy-weighted sum rule.

\item[Conclusions]
Despite a reduction of Coulomb relative to nuclear cross sections in light nuclei with respect to previous analyses of the same kind in medium-mass and heavy nuclei, forward-angle proton scattering at energies of several hundred MeV remains a valuable tool to extract the nuclear $E1$ response.
However, there is a larger model dependence, in particular for excitation energies above 20 MeV, due to the need to constrain the continuum background with additional assumptions. 
The overall success of the shell-model approach to describe the features of the experimental photoabsorption cross sections motivates its application in large-scale reaction network calculations aiming at an understanding of the mass composition of ultrahigh-energy cosmic rays.

\end{description}
\end{abstract}


\maketitle


\section{\label{sec:1}Introduction}

The electric dipole response of nuclei is a key observable for an understanding of nuclear structure and for many applications \cite{bracco19,zilges22}.
It is dominated by the IVGDR, which has been a subject of experimental investigations since its discovery more than 90 years ago. 
A large part of the information in heavy nuclei stems from the $(\gamma,xn)$ reaction measured with monochromatic photon beams \cite{zilges22,berman75}.
There is also interest in the phenomenon of a resonance-like structure below the neutron threshold in nuclei with neutron excess commonly termed pygmy dipole resonance \cite{bracco19,zilges22,savran13}.
Its nature is subject of debate \cite{lanza23,repko19,vonneumanncosel24}. 
Alternatively, relativistic Coulomb excitation in inelastic proton scattering at very forward angles has been established as a tool to study the $E1$ strength in a single experiment from low excitation energies across the IVGDR \cite{vonneumanncosel19}.
A series of nuclei has been studied with this technique ranging from $^{208}$Pb to $^{40}$Ca \cite{tamii11,bassauer20,martin17,brandherm24,brandherm25,birkhan17,fearick23} with the goal to extract information on the neutron skin thickness and symmetry energy properties \cite{vonneumanncosel25}.

While the majority of experimental work has focused on heavy nuclei, there is renewed interest in $E1$ strength distributions of light $p$- and $sd$-shell nuclei in the framework of the PANDORA project \cite{tamii23}, which aims at an understanding of the mass distribution of ultrahigh-energy cosmic radiation (UHECR) \cite{anchordoqui19}.
UHECR is of extragalactic origin and its interaction along the travel path is dominated by photoabsorption of cosmic background radiation boosted to the IVGDR energy region in the center-of-mass system.
Thus, systematic knowledge of the photoabsorption in light nuclei and their subsequent particle decay is required. 
Since it will not be possible to study all relevant reactions, the PANDORA project aims at a measurement of photoabsorption cross sections and the coincident particle decay in selected nuclei serving as benchmark for theoretical models of the full reaction network.
While such measurements with monoenergetic real photon beams are part of the program of the ELI-NP facility \cite{ELI-NP} once it is completed, at present the above-mentioned $(p,p^\prime)$ scattering at forward angles \cite{vonneumanncosel19,neveling11} can be employed for such experiments. 

Traditionally, the theoretical description of the IVGDR is a domain of  density functional theory \cite{bender03}, since the single-particle model spaces spanning several oscillator shells are prohibitive for shell-model (SM) approaches.
However, recent progress allows SM calculations in $sd$-shell nuclei with sufficiently large models spaces to describe the IVGDR \cite{sieja23,lenoan25}. 
Experimental photoabsorption data in $sd$-shell nuclei show large discrepancies with each other in many cases.
Thus, new results with a different experimental technique are of considerable interest.
The present work extends the approach described in Ref.~\cite{fearick23} for $^{40}$Ca to $sd$-shell nuclei, most of them originally measured for the study of isoscalar spin-$M1$ transitions \cite{matsubara15}.

The extraction of the $E1$ cross sections originating from Coulomb excitation through a multipole decomposition analysis (MDA) becomes increasingly difficult in lighter nuclei, since the Coulomb cross sections scale with target charge while the nuclear background is approximately independent of mass or charge.
However, the successful analysis for the case of $^{40}$Ca \cite{fearick23} motivates an extension to even lighter nuclei. 
The IVGDR is known to be very fragmented in the $sd$ shell and extends to very high excitation energies \cite{eramzhyan86}.
Its fine structure has been studied in Ref.~\cite{fearick18}.
The limited momentum acceptance of the magnetic spectrometer used to measure the scattered protons in the $(p,p^\prime$) experiment at $0^\circ$ does not permit to extract the dipole polarizability, which would be of interest to decompose volume and surface contributions of the symmetry energy \cite{vonneumanncosel16}. 
Rather, the data serve as a global test of shell-model calculations of the $E1$ response across the $sd$ shell.
Besides the $N = Z$ nuclei $^{24}$Mg, $^{28}$Si, $^{32}$S, and $^{36}$Ar studied in Ref.~\cite{matsubara15},  experimental results are available for $^{20}$Ne and $^{26}$Mg.
We also include a reanalysis of $^{40}$Ca to provide an estimate of uncertainties due to the assumptions of the present analysis described below by comparison with the results presented in Ref.~\cite{fearick23}. 

The paper is organized as follows.
Section \ref{sec:2} provides details of the experiment, of the MDA approach to extract $E1$ cross sections, and of the conversion to photoabsorption cross sections.
The comparison with previous work and predictions of an artificial neural network (ANN) application is discussed in Sec.~\ref{sec:3}.
Section \ref{sec:4} describes the shell-model calculations and the comparison to the present results.
Conclusions are given in Sec.~\ref{sec:5}.

\section{\label{sec:2}Experiment and data analysis}

\subsection{\label{sec:2A} Details of the experiment at RCNP}

Experimental data were acquired using a 295 MeV proton beam at the RCNP cyclotron facility in Osaka, Japan. 
Inelastic $(p,p')$ energy spectra were measured  for a series of nuclei using the Grand Raiden (GR) magnetic spectrometer \cite{fujiwara99}. 
The setup permits measurements at forward scattering angles  $0^\circ$,$2.5^\circ$, and $4.5^\circ$ \cite{tamii09}.
The large angular acceptance allows to extract data for several angles at a given GR angle by software cuts.
Data were recorded at scattering angles (in the laboratory frame)  $0.40^\circ$, $1.0^\circ$, $1.74^\circ$  (GR at $0^\circ$), $2.38^\circ$ and $3.18^\circ$ (GR at $2.5^\circ$), $4.39^\circ$ and $5.15^\circ$ (GR at $4.5^\circ$), and  $6.2^\circ$, $8.2^\circ$, $10.1^\circ$, $12.0^\circ$, and $14.0^\circ$ with the GR spectrometer at those angles.

\begin{table}
\caption{\label{tab:target} Target properties.}
\begin{tabular}{llrr}
\hline
\hline
Nucleus & State & Areal density & Enrichment \\
& & (mg/cm$^2$) & (\%) \\
\hline
$^{20}$Ne & gas & 1.06(3) & 100.0 \\
$^{24}$Mg & foil & 2.50(3) & 100.0 \\
$^{26}$Mg & foil & 1.55(5) & 100.0\\
$^{28}$Si & foil & 2.29(2) & 92.2 \\
$^{32}$S & sheet & 23.5(5) & 95.0 \\
$^{36}$Ar & gas & 1.04(3) & 100.0 \\
$^{40}$Ca & foil & 2.98(3) & 100.0 \\
\hline
\hline
\end{tabular}
\end{table}

Target nuclei were $^{20}$Ne, $^{24}$Mg,
$^{26}$Mg, $^{28}$Si, $^{32}$S, $^{36}$Ar and $^{40}$Ca.
Details of the target compositions are given in Table~\ref{tab:target}. 
Solid targets were self-supporting foils.
The setup for gaseous targets is described in Ref.~\cite{matsubara12}. 
In order to remove scattering from the aramide windows of the gas target, spectra obtained with no gas present were subtracted to give the net gas response. 
The production of self-supporting sulphur targets is described in Ref.~\cite{matsubara09}.

The spectrometer was dispersion matched to provide an energy resolution of typically 20--30 keV full width at half maximum. 
The settings of the GR permitted excitation energies up to 25 MeV. 
The low excitation energy region was cut off at 6-8 MeV, depending on the target, due to the geometrical limitations of the $0^\circ$ setup.
The raw data analysis is described in Ref.~\cite{tamii09}.

\begin{figure}
    \centering
    \includegraphics[width=\columnwidth]{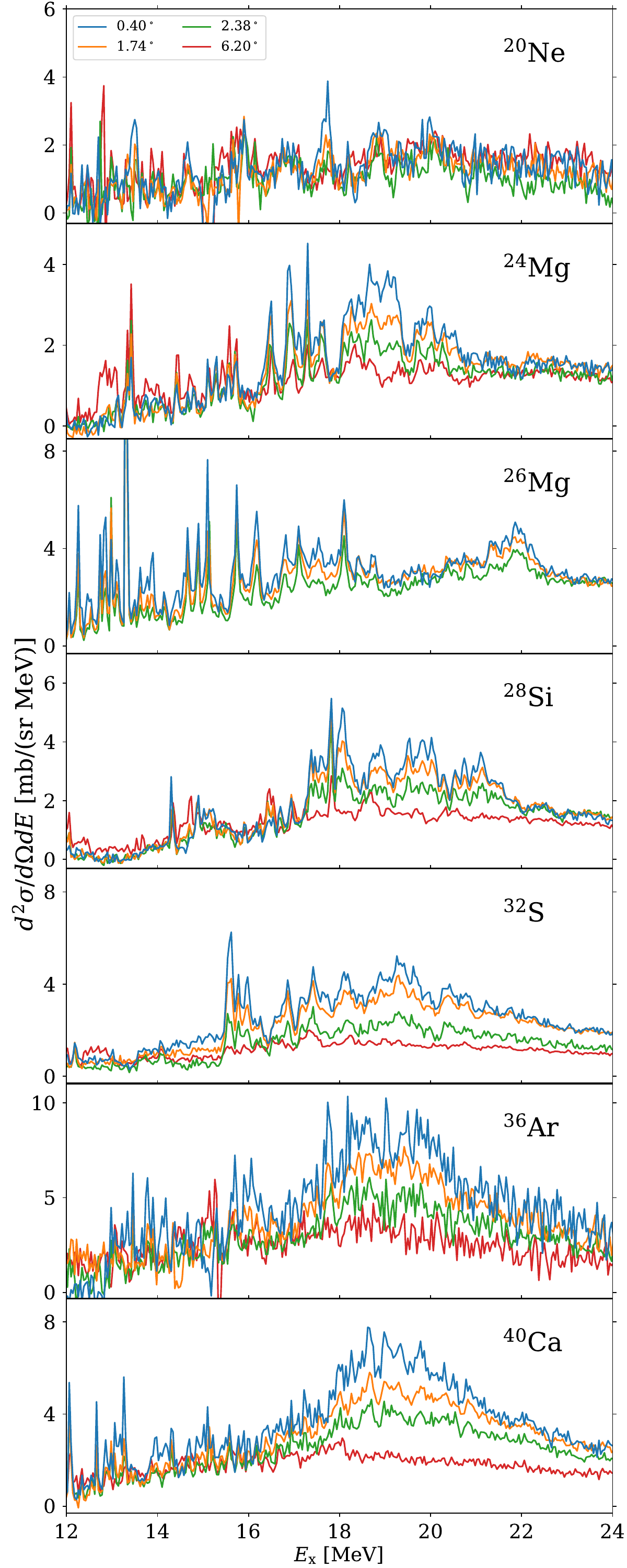}
    \caption{Differential cross sections for inelastic scattering of 295 MeV protons for various $sd$-shell nuclei in the energy region of the IVGDR. 
    Spectra are shown for laboratory scattering angles of $0.40^\circ$, $1.74^\circ$, $2.38^\circ$, and $6.20^\circ$. 
    }
    \label{fig:allspec}
\end{figure}

Representative spectra of all targets are shown in Fig.~\ref{fig:allspec} at angles $0.40^\circ$, $1.74^\circ$, $2.38^\circ$, and $6.20^\circ$ for an excitation energy range $12 - 24$ MeV.
For $^{26}$Mg, only data for scattering angles up to $5.15^\circ$ are available.
At lower excitation energies the spectra are dominated by resolved transitions, which are mostly of $M1$ character \cite{matsubara15}, and the few candidates for $E1$ transitions have very small cross sections \cite{matsubara10}.

\subsection{\label{sec:2B} Multipole decomposition analysis}

For further analysis and to improve statistics, the data were rebinned to 250 keV, covering the excitation energy range 10-25 MeV.
For each nucleus in the data set a multipole decomposition analysis (MDA) was performed on the rebinned data, using theoretical collective form factors calculated as derivative of the optical potential using the distorted-wave Born approximation code DWUCK4 \cite{DWUCK}.
Optical model parameters were available from measurements of elastic scattering \cite{matsubara10}.

\begin{figure}
    \centering
    \includegraphics[width=\columnwidth]{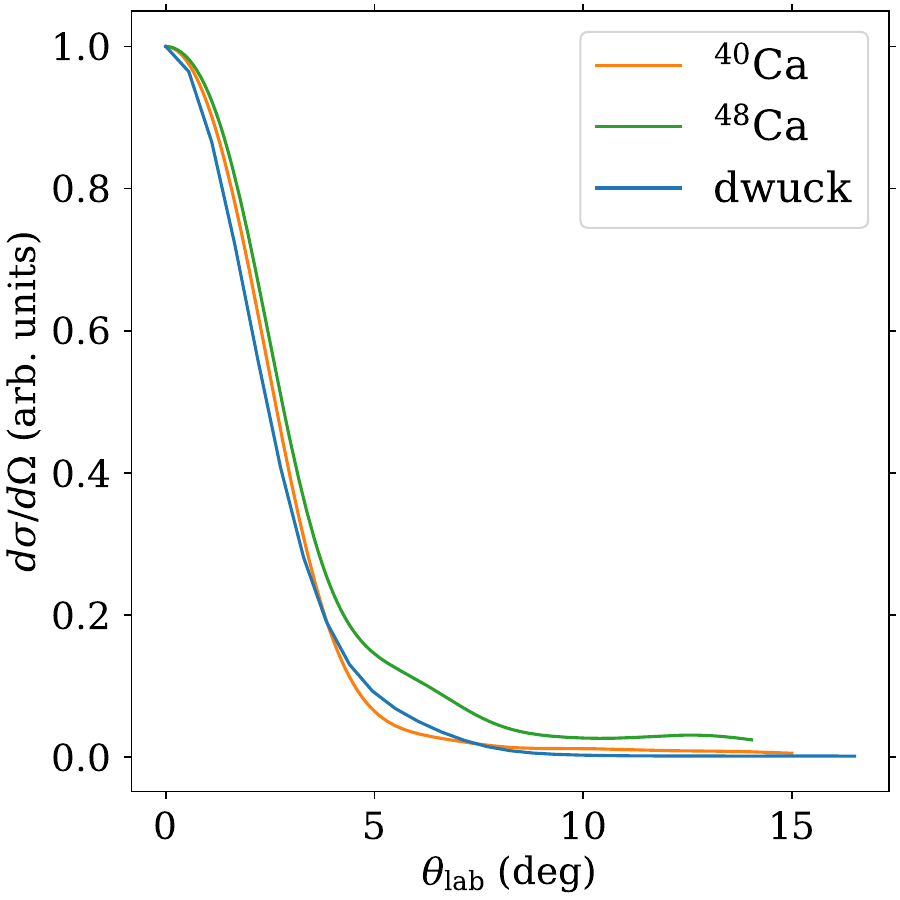}
    \caption{Comparison of $E1$ excitation cross section for an excitation energy of 18 MeV as calculated by DWUCK4 (blue) and from microscopic calculations for $^{40}$Ca (orange) and $^{48}$Ca, scaled to $^{40}$Ca kinematics (green). 
    }
    \label{fig:comp-dwba}
\end{figure}

In order to test the impact of the collective approximation, a detailed comparison was made for $^{40}$Ca with the analysis described in Ref.~\cite{fearick23} based on microscopically calculated angular distributions.
For multipoles $L = 0$ and $L > 1$ little difference was observed.
The angular distributions of $E1$ cross sections are more critical because of the interference of Coulomb and nuclear contributions \cite{vonneumanncosel19}. 
The corresponding angular distributions are compared in Fig.~\ref{fig:comp-dwba} also including the microscopic prediction for $^{48}$Ca \cite{birkhan17} scaled to $^{40}$Ca kinematics. 
The agreement was judged adequate for the purposes of this analysis.

A major problem in the analysis is the large background that arises from preequilibrium events in the scattering dominated by quasifree scattering. 
Two approaches were considered. 
In the first, the angular distribution for these continuum processes was taken from the empirical systematics of Ref.~\cite{kalbach88}. 
This was used in the linear fit of the MDA together with the DWUCK angular distributions for $1^-$, $2^+$, and $3^-$ multipoles. 
In the second approach, based on experience in the previous analysis of the $^{40}$Ca data \cite{fearick23}, a fixed shape in energy was used to further constrain the continuum contribution. 
This latter approach was also used in the present study.

\begin{figure}
    \centering
    \includegraphics[width=\columnwidth]{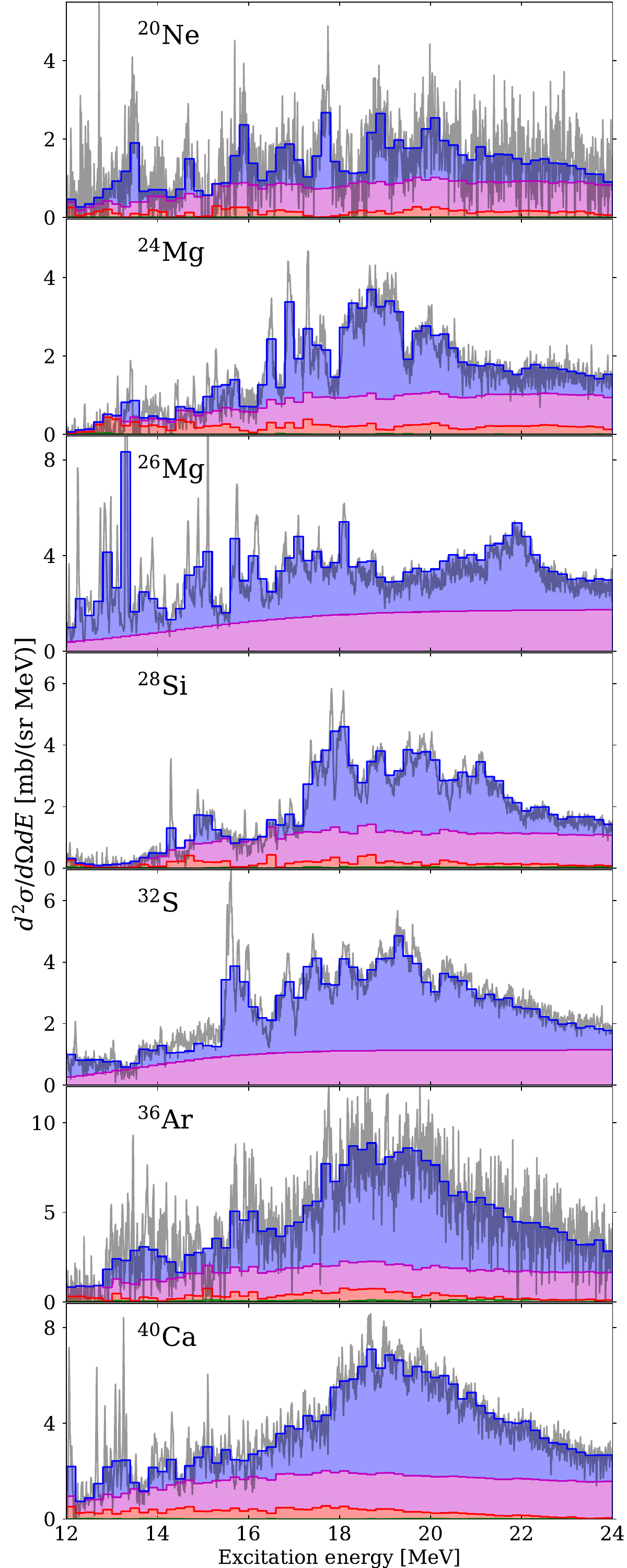}
    \caption{Results of the multipole decomposition of the cross section for the spectra measured at a laboratory scattering angle of $0.4^\circ$. The blue region shows the contribution from $E1$, the orange region the contribution from $E2$, and the magenta region the contribution from quasifree scattering. $E3$ is negligible.}
    \label{fig:allmda}
\end{figure}

The resulting decomposition of cross sections is shown for all nuclei in Fig.~\ref{fig:allmda}, for the spectra measured at $0.4^\circ$. 
The most important contributions are $E1$ and quasifree scattering. 
One observes a general increase of the $E1$ cross sections with target charge, while the quasifree cross of the order 1-2 mb/(sr MeV) show little mass or charge dependence.
$E2$ contributions are small and higher multipoles are negligible at this very forward angle.

\subsection{\label{sec:2C} Extraction of electric dipole strength}

The $E1$ cross sections resulting from the MDA were converted to an effective photoabsorption cross section using the virtual photon method.
The virtual photon cross sections were calculated in the eikonal approach \cite{bertulani93}. 
These cross sections rise steeply in the small-angle region. 
Since the solid angle was defined by a gate on the focal plane detector, the probability density function of the scattering angle was calculated by numerical integration over the gate area and used to calculate the average virtual photon cross section for a given scattering angle, as described e.g.\ in Ref.~\cite{bassauer20}.

\section{\label{sec:3} Results}

\subsection{\label{sec:3A} Comparison to previous work}

\begin{figure}
    \centering
    \includegraphics[width=\columnwidth]{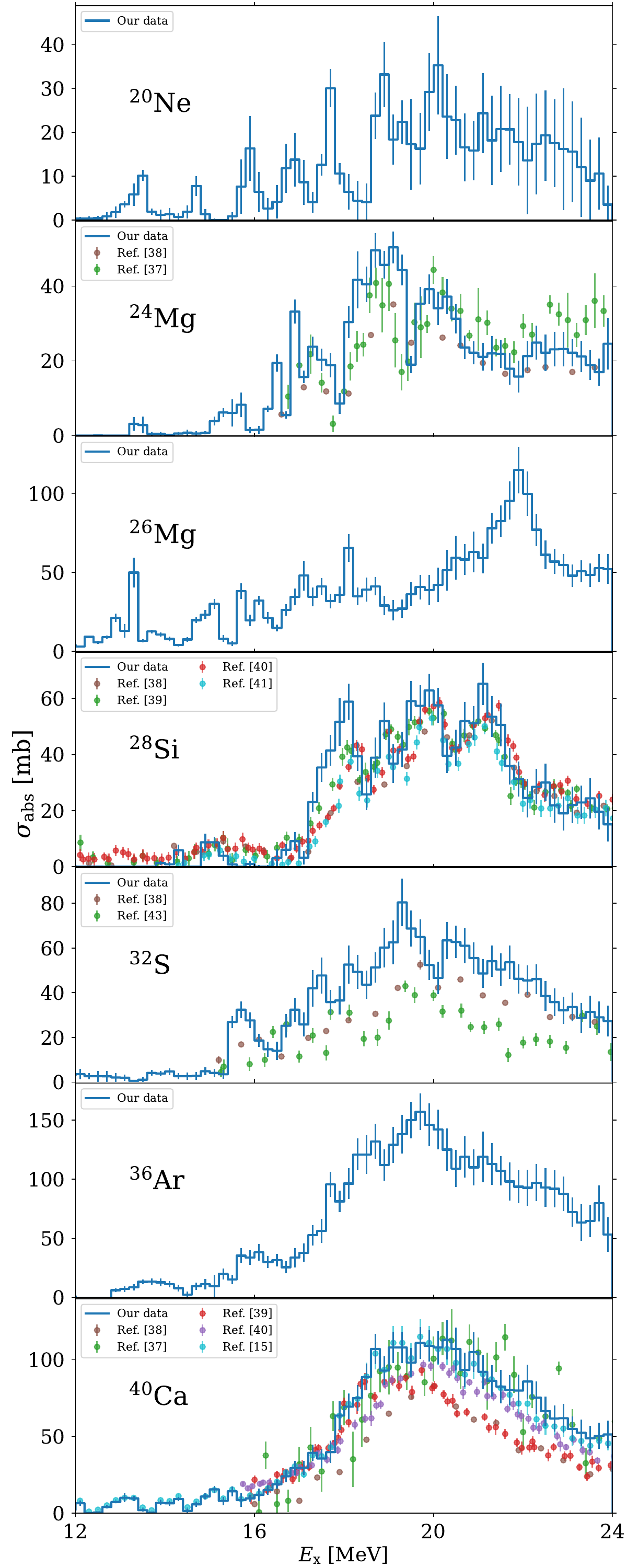}
    \caption{Photoabsorption cross sections obtained from the $E1$ $(p,p^\prime)$ cross sections using the virtual photon method.
    The blue histograms show the data obtained in this experiment, while the points show previous results from the IAEA photon strength function database with the references given in the text.}
    \label{fig:all_sigabs}
\end{figure}

The resulting absorption cross sections are shown in Fig.~\ref{fig:all_sigabs} together with previous experimental data extracted from the latest version of the IAEA photon strength function database \cite{kawano20}.
In general, the lighter nuclei up to $^{28}$Si show pronounced structure, while for heavier nuclei in the $sd$ shell a resonance-like structure starts to emerge.
However, it should be noted, that significant isospin splitting is observed throughout the $sd$ shell \cite{eramzhyan86}, and  a large part of the $T_> = T_{\rm g.s.} + 1$ component typically lies outside the studied energy range.

For $^{20}$Ne, $^{26}$Mg, and $^{36}$Ar no previous measurements have been reported.
In $^{24}$Mg, the results of Ref.~\cite{dolbilkin66} (green circles) qualitatively agree with the main structures seen in the present data above 16 MeV, but find smaller cross section values below 18 MeV, approximately equal ones between 18 and 22 and larger ones at higher excitation energies. 
The data of Ref.~\cite{wyckoff65} (brown circles) find the same structures but overall smaller cross sections.
The $^{28}$Si cross sections derived in the present analysis agree on the position of main structures and absolute magnitudes with previous work \cite{wyckoff65} (brown circles),\cite{bezic68} (green circles), \cite{ahrens75} (red circles), and \cite{schelhaas91} (cyan circles).
The only exception is the peak at about 18 MeV, where the present work indicates larger cross sections.
A high-resolution study of the $^{28}$Si$(\gamma, {\rm abs})$ reaction in the excitation region 17.5-21.5 MeV \cite{harada01} is in good agreement but shows more fine structure in the observed peaks averaged out in the present analysis by the 250 keV binning. 
The results of Ref.~\cite{harada01} also show a larger cross section around 18 MeV than previous work in line with our present findings.

For $^{32}$S,two previous measurements exist by Ref.~\cite{wyckoff65} (brown circles) and Ref.~\cite{dolbilkin68} (green circles). 
Both have limited energy resolution and do not agree with each other on cross section magnitudes.
The present results show considerably more structure and generally larger cross sections.
Finally, the $^{40}$Ca result agrees closely with the MDA analysis of the same data by Ref.~\cite{fearick23} based on microscopically calculated angular distributions and an effective proton-nucleus interaction \cite{love81,franey85}.
It also agrees well with the total photoabsorption measurement of Ahrens {\it et al.}~\cite{ahrens75}, while older measurements agree on the main structure but find overall smaller \cite{wyckoff65} (brown circles), \cite{bezic68} (red circles) or larger \cite{dolbilkin66} (green circles) cross sections.

\subsection{\label{sec:3B} Comparison to ANN predictions}

\begin{figure}
    \centering
    \includegraphics[width=0.99\columnwidth]{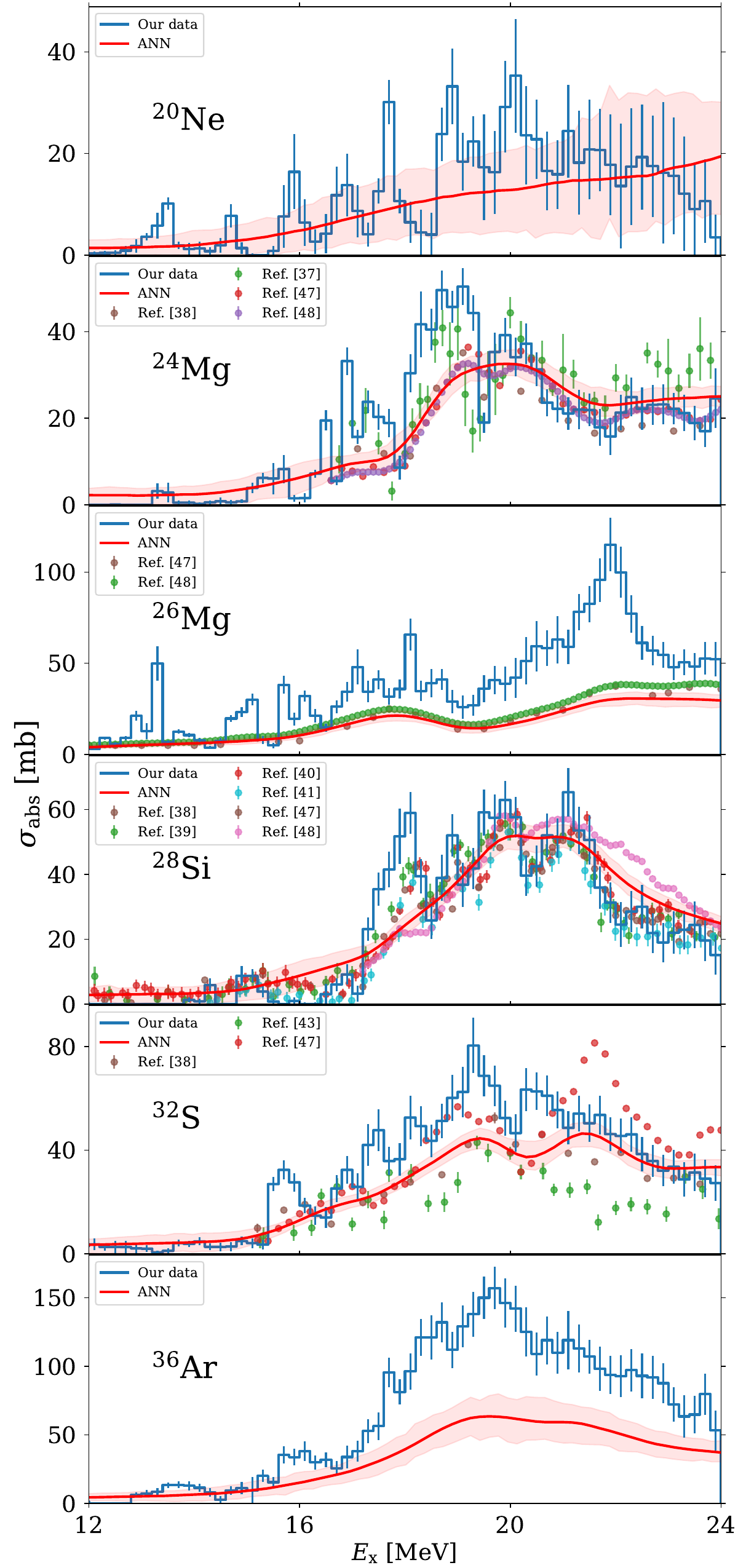}
    \caption{Comparison of the present (histogram) and previous (data points) photoabsorption measurements for $^{20}$Ne, $^{24,26}$Mg, $^{28}$Si, $^{32}$S, and $^{36}$Ar with data-driven predictions of dipole strength functions based on ANN (red lines with shaded uncertainty band).}
    \label{fig:all_sigabs_ANN}
\end{figure}

Recently, a data-driven analysis of dipole strength functions using ANN has been presented \cite{jiang25}.
It was shown that reliable predictions can be derived for heavy nuclei, but also down to lighter nuclei like $^{40,48}$Ca.
This work allows a test for even lighter nuclei in the $sd$ shell.
The present data are compared in Fig.~\ref{fig:all_sigabs_ANN} with the ANN results shown as red lines with shaded uncertainty bands estimated as described in Ref.~\cite{jiang25}.
For reference, we also show the previous measurements discussed in Sec.~\ref{sec:3A}.
Additionally, the results of Refs.~\cite{ishkanov2002,varlamov2003} are included, since they entered into the data base of the ANN analysis.
These are mostly no direct measurements of total photoabsorption but compilations based on summing partial cross sections.

The comparison shows a mixed success of the ANN predictions.
In $^{20}$Ne, the average energy dependence is well reproduced.
This also holds for $^{24}$Mg at excitation energies above 19 MeV, while the prominent structures around 17 and 19 MeV are missed.
The data in $^{28}$Si are also reasonably well described except for the prominent peak around 18 MeV.
On the other hand, results in $^{26}$Mg and $^{36}$Ar are generally underestimated.
Below 22 MeV, this is also true for $^{32}$S.

Not unexpectedly, the ANN predictions are not capable of reproducing the strong fragmentation of the IVGDR observed in the $sd$ shell.
The ANN training data are mostly from heavier nuclei, which show compact resonances.
As pointed out above, total photoabsorption data in light nuclei are scarce.
Except for $^{20}$Ne and $^{36}$Ar, where no previous results exist, the ANN predictions closely follow an average of the data available prior to the present results, cf.\ $^{24,26}$Mg as examples.
It would be interesting to see if the inclusion of high-resolution results like the present work or results from the PANDORA project \cite{tamii23} in the near future improve ANN predictions in these light nuclei overall as well as of details of the energy dependence. 

\section{\label{sec:4} Comparison to shell-model calculations}

\subsection{\label{sec:4A} Details of the shell-model approach}
Our approach to treating $E1$ strength of $sd$-shell nuclei within the Configuration Interaction Shell-Model framework (CI-SM) with the PSDPF interaction \cite{bouhelal11}, was recently presented in detail in Refs.~\cite{lenoan25, le_noan_configuration_2025}. 
A systematic comparison of our results to other available theoretical models can be found in Ref. \cite{lenoan_UHECR}.

We remind that the shell-model Hamiltonian reads
\begin{equation}
H=\sum_i \epsilon_i c_i^\dagger c_i+ \sum_{ijkl}V_{ijkl}c_i^\dagger c_j^\dagger c_l c_k+\beta H_{\textrm{c.m.}},
\end{equation}
where the center-of-mass (c.m.) Hamiltonian, scaled by a coefficient $\beta = 10$, is included to shift the spurious c.m.\ eigenvalues into an energy range outside the scope of the present study \cite{gloeckner_spurious_1974}. 
We allow only for $0\hbar \omega$ ground states (GS) and $1\hbar \omega$ excited states.
The $E1$ strength is generated by the isovector dipole operator:
\begin{equation}
\hat O_{1\mu}=-e\frac{Z}{A}\sum_{i=1}^N r_i Y_{1\mu}(\hat r_i)+e\frac{N}{A}\sum_{i=1}^Z r_iY_{1\mu}(\hat r_i).
\label{oper}
\end{equation}
As discussed in \cite{lenoan25}, the $E1$ operator must be renormalized, or “dressed,” to incorporate correlations beyond the model space and we determined the $E1$ effective charges for $sd$-shell nuclei as $(e_{\textrm{eff}}^n, e_{\textrm{eff}}^p) = (-0.8Z/A, 0.8N/A)$, corresponding to an overall reduction of the total strength by a factor of $0.64$.
The reduced transition probability in CI-SM is further calculated as 
\begin{equation}
B_{\nu 0}=\frac{1}{2J_0+1}\langle \nu||\hat O|| 0\rangle^2,
\end{equation}
where $|{0}\rangle$ is the nuclear GS, $|{\nu}\rangle$ an excited eigenstate, $J_0$ the GS total angular momentum quantum number and $\hat O$ the transition operator, in our case from Eq.~(\ref{oper}). 
The $B(E1)$ distributions are obtained through the Lanczos strength-function method, enabling an efficient determination of the strength per energy interval \cite{caurier05}. 
The computations of this work were performed using the $m$-scheme shell-model code ANTOINE \cite{caurier05} with 300 Lanczos iterations, which ensures the convergence of the results. For comparison to data, the discrete $B(E1)$ distributions were additionally convoluted with Lorentzians
and converted to photoabsorption cross sections:
\begin{equation}
\sigma_{abs}(E)=\frac{16\pi^3 e^2}{9(\hbar c)}\sum_{\nu}\frac{1}{\pi}\frac{\Gamma/2}{(E-E_\textrm{x}(\nu))^2+(\Gamma/2)^2}E_\textrm{x}(\nu)B_{\nu 0},
\label{Eq-SE1}
\end{equation}
where $E_x(\nu)$ is excitation energy of the $|\nu\rangle$ eigenstate and $\Gamma=0.3$ or $1$ MeV, respectively.

We also consider the nuclear dipole polarizability\cite{lipparini_sum_1989, Ring80}:  
\begin{equation}
\alpha_D=\frac{8\pi}{9}S_{-1}
\end{equation}
where $S_{-1}$ is the $k=-1$ moment of the distribution
\begin{equation}
S_k=\sum_\nu(E_\nu-E_0)^k |\langle\nu|\hat O|0\rangle|^2. 
\label{S}
\end{equation}
The centroid and width read
\begin{eqnarray}
\bar S=\frac{S_1}{S_0},\quad \Delta S=\sqrt{\frac{S_2}{S_0}-\bar S^2},
\label{eq-moments}
\end{eqnarray}
with $S_k$ defined in Eq.~(\ref{S}).

\subsection{\label{sec:4B} Discussion}

\begin{figure*}
    \centering
    \includegraphics[width=0.9\textwidth]{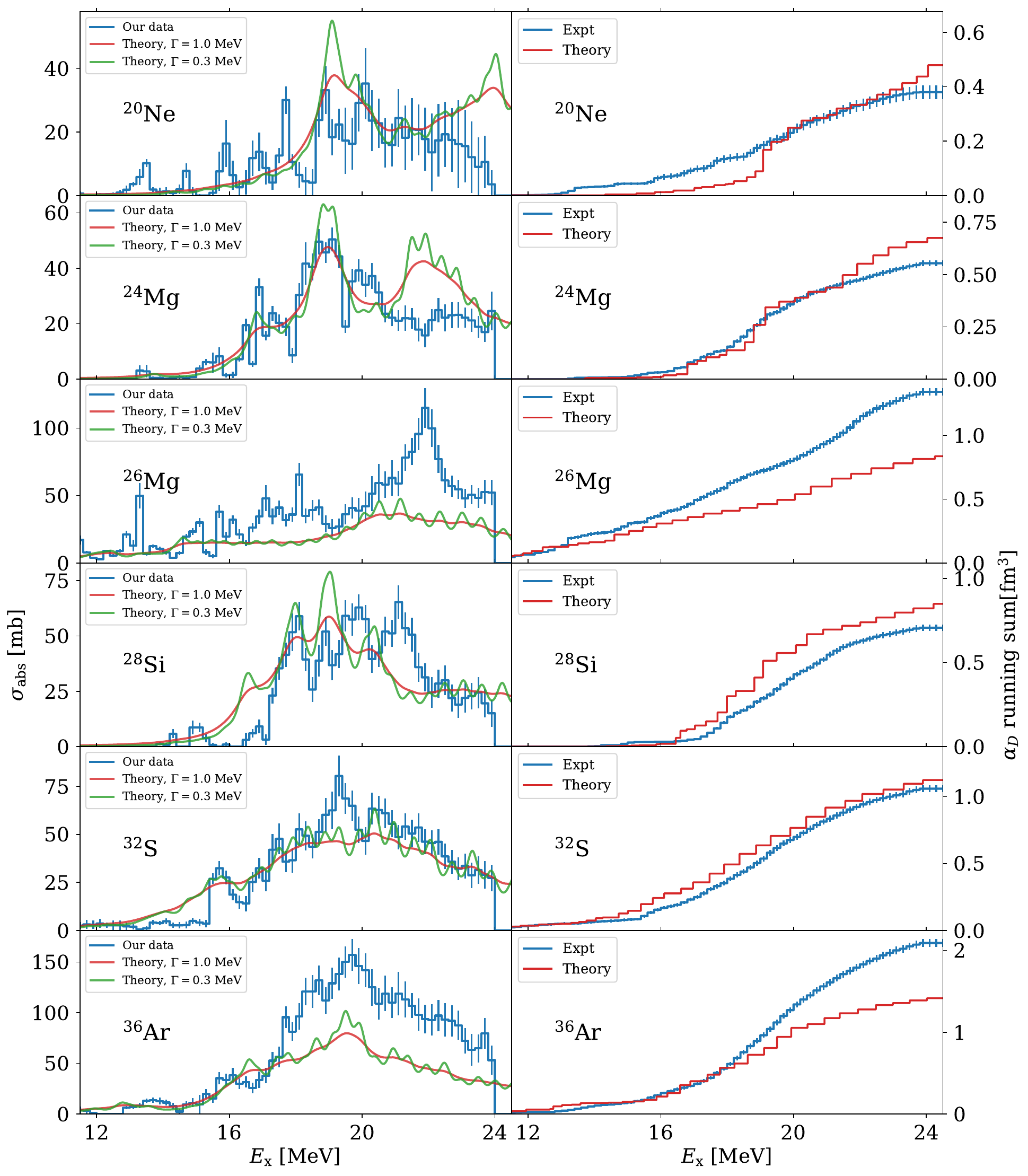}
    \caption{
    Left panel: Experimental photoabsorption cross sections (blue histograms) together with the shell-model predictions. 
    The theoretical data are convoluted with two Lorentzian curves for comparison of gross and fine structures with full widths at half maximum $\Gamma=1.0$ (red curves) and 0.3 MeV (green curves), respectively. 
    Right panel: Experimental(blue histogram) and theoretical (red histograms) running sums of the dipole polarizability, $\alpha_D$, as  a function of excitation energy.
    The theoretical sums are extracted from the unconvoluted results.}
    \label{fig:shellmodel}
\end{figure*}

Figure \ref{fig:shellmodel} summarizes the experimental and theoretical photoabsorption cross sections for the nuclei studied in the present work.
The left column displays the dipole strength distributions while the right column shows the running sums of $\alpha_D$.
The latter do not show saturation in the studied excitation energy window indicating that significant $E1$ strength lies at excitation energies above 24 MeV.

A test of gross features of the measured strength distributions is provided by calculations of the centroids and widths, Eq.~(\ref{eq-moments}). 
Because of the potentially large systematic uncertainties of the data above 20 MeV for some of the nuclei, the comparison of theoretical and experimental centroids is restricted to the excitation energy region $12 -20$ MeV. 
The results are summarized in Tab. \ref{tab-CISM}.
As discussed in earlier work \cite{le_noan_configuration_2025}, the present shell-model calculations reproduced the centroids of 26 experimentally known $sd$-shell nuclei with a root-mean-square (rms) deviation of $0.84$ MeV, however systematically at too low energy when compared to available photoabsorption data. 
The present photoabsorption measurements seem to confirm this deficiency of shell-model predictions and 
additionally suggest a mass-dependent deviation within this energy range. 
Nevertheless, across the six measurements reported here, the rms deviations of the centroids and widths are $0.53$~MeV and $0.28$~MeV, respectively, indicating an overall good agreement.

\begin{table}
\caption{Centroids and widths (in MeV), defined in Eq. \ref{eq-moments}, calculated between $12$ and $20$ MeV.\label{tab-CISM}}
\begin{tabular}{ccccccc} 
\hline 
\hline 
\addlinespace[1ex]
   & {$\overline{S}$} & {$\overline{S}_\textrm{exp}$} & {$\overline{S} - \overline{S}_\textrm{exp}$ } &  {$\Delta S$} & {$\Delta S_\textrm{exp}$}  & {$\Delta S - \Delta S_\textrm{exp}$ } \\ 
\addlinespace[1ex]
\hline
\addlinespace[1ex]
     {$^{20}$Ne} & {$18.40$} & {$17.50$} & {$0.90$} & {$1.40$} & {$1.99$} & {$-0.59$} \\
     {$^{24}$Mg} & {$18.06$} & {$18.15$} & {$-0.09$} & {$1.43$} & {$1.34$} & { $ 0.09$} \\
     {$^{26}$Mg} & {$16.49$} & {$16.65$} & {$-0.16$} & {$2.22$} & {$2.15$} & { $ 0.06$} \\
     {$^{28}$Si} & {$17.95$} & {$18.41$} & {$-0.45$} & {$1.38$} & {$1.22$} & { $ 0.23$} \\
     {$^{32}$S} & {$17.23$} & {$17.82$} & {$-0.59$} & {$1.86$} & {$1.65$} & { $ 0.21$} \\
     {$^{36}$Ar} & {$17.49$} & {$18.01$} & {$-0.52$} & {$1.87$} & {$1.64$} & { $ 0.23$} \\
\hline
\hline 
\end{tabular}
\end{table}

Below $18$ MeV, the CI-SM cross section is almost always below the measured one. Note that in $N=Z$ nuclei the $E1$ transitions generated by the isovector operator Eq.~(\ref{oper}) between the states of the same isospin are forbidden by isospin symmetry. The missing strength at the lowest energies in $N=Z$ nuclei might be due to the isoscalar part of $E1$ response \cite{graef77,friedrich89,campos95} which is not accounted for in the present CI-SM calculations. 

A detailed comparison between experimental and theoretical photoabsorption cross sections is shown in the left column of Fig.~\ref{fig:shellmodel}.
Starting with $^{20}$Ne, the CI-SM prediction exhibits a two-peak structure reflecting its deformation. 
While the first peak around $18–20$ MeV is well reproduced, the experimental data deviate from the CI-SM results above $\sim 22$  MeV. In this energy region, the experimental photoabsorption cross section shows a steep decrease, nearly vanishing at 24 MeV. This trend is not confirmed by theory nor the ANN predictions. This could indicate that the background subtraction above 20 MeV was not fully optimal, possibly due to the reduced signal-to-background ratio, cf.\ Fig. \ref{fig:allmda}. 

In $^{24}$Mg we observe the best overall agreement between theory and experiment up to $20$ MeV, with the fragmentation of the cross section being also well reproduced. This results in a relative difference between the theoretical and experimental centroid values of 0.09 MeV only, with a similarly small discrepancy for the width. In this energy region, earlier measurements are also in better agreement with the present experimental result than for other nuclei. Above 20 MeV, however, the CI-SM predicts another peak, whereas the measured cross section remains relatively flat. 

A satisfactory agreement between theory and experiment is also found in $^{28}$Si and $^{32}$S. 
In $^{28}$Si the CI-SM results exhibit the same fragmentation as the experimental data, with three distinct peaks, though the 
theoretical predictions seems shifted towards lower energies, leading to an overestimation of the computed polarizability along the whole energy range covered by our data. A slight shift of the computed strength towards higher energies would bring considerable improvement. In $^{32}$S, experimental data and theoretical predictions closely follow each other over most of the energy range, except in the region of the IVGDR peak, where the CI-SM cross sections lie slightly below the measured values. 

In $^{26}$Mg, an overall underestimation of the cross section by theory with respect to the experimental data is observed. Nevertheless, the two-peak structure is well reproduced, and the centroid energy and width, being quantities normalized to the total strength, remain in good agreement with the measured values. Previous measurements and the ANN prediction are closer to theory than to the present experimental data. Note that the effective $E1$ operator was chosen in Ref. \cite{lenoan25} for this region such that the overall CI-SM strength is consistent with the Thomas-Reiche-Kuhn (TRK) sum rule within a $~15\%$ enhancement. For the CI-SM prediction to reproduce the present measurement, an enhancement factor of about $100\%$ would be required. Such a large value appears difficult to justify, as enhancement of this magnitude, in this mass region, have not been reported so far in the GDR energy range. Moreover, the enhancement factor is generally expected to vary smoothly with mass, suggesting that such a sudden increase is unlikely to have a physical origin.

The same discrepancy is observed in $^{36}$Ar beyond 18 MeV, where the theory considerably underestimates theexperimental data, although the shape of the cross section and its fragmentation are similar. Proceeding in the same way as for the $^{26}$Mg, reproducing the present experimental cross section would require an unlikely enhancement factor of about $100\%$. We note additionally that the ANN predictions in those two cases are found to be in good agreement with the theoretical cross section.


\section{\label{sec:5} Conclusions}

To summarize, we present new photoabsorption data for the $sd$-shell $N=Z$ nuclei and $^{26}$Mg.
These were derived from forward-angle inelastic proton scattering data at 295 MeV based on a MDA and subsequent conversion of the $E1$ Coulomb excitation to photoabsorption cross sections, as described e.g.\ in Ref.~\cite{vonneumanncosel19}.
The MDA shows that the main background contributions stem from continuum cross sections dominated by knockout processes.
The need to constrain these contributions with additional assumptions makes the results more model-dependent than in heavy nuclei, in particular above excitation energies of about 20 MeV.
Fair agreement with previous experiments is observed for $^{24}$Mg and $^{28}$Si, while results disagree for $^{32}$S.
No prior data were reported for $^{20}$Ne, $^{26}$Mg, and $^{36}$Ar.

The results are compared to data-driven ANN predictions \cite{jiang25}.
For light nuclei, these closely follow the scarce previously available data in this mass region. 
It is also clear that because of the lack of sufficient training data detailed structures of the $E1$ strength distributions as observed in the present data cannot be reproduced.  
Overall fair agreement is observed for $^{20}$Ne, $^{24}$Mg, and $^{28}$Si.
For the other nuclei, the ANN predictions are significantly smaller than the present cross sections. 

With the exception of the magnitudes of cross sections in $^{26}$Mg and $^{36}$Ar, recent CI-SM calculations \cite{lenoan25,le_noan_configuration_2025} provide an overall satisfactory qualitative and quantitative description of the data. 
This encourages application of such results in large-scale reaction network calculations aiming at an understanding of the mass composition of UHECR \cite{lenoan_UHECR} within the PANDORA project \cite{tamii23}.
In $^{26}$Mg and $^{36}$Ar, the present photoabsorption cross sections are much larger than the CI-SM results normalized to an enhancement of the TRK sum rule expected in this mass region. 
This problem needs further experimental investigation.

\begin{acknowledgments}
The accelerator group at RCNP is thanked for high-quality dispersion-matched beams for the experiments. 
We are indebted to W.~G.~Jiang for providing us with the ANN results.
This work was supported by the Deutsche Forschungsgemeinschaft (DFG, German Research Foundation) under Contract No.\ SFB 1245 (Project ID No.\ 79384907).
RWF acknowledges support of the University of Cape Town Science Faculty Research Committee.
OLN and KS acknowledge support of the Interdisciplinary Thematic Institute QMat, as part of the ITI 2021-2028 program of the University of Strasbourg, CNRS and Inserm, supported by IdEx Unistra (ANR 10 IDEX 0002), and by SFRI STRAT’US project (ANR 20 SFRI 0012) and EUR QMAT ANR-17-EURE-0024 under the framework of the French Investments for the Future Program.
\end{acknowledgments}

\bibliography{E1_sd-shell}

\end{document}